\documentclass[showpacs,preprintnumbers,amsmath,reprint,aps,prl]{revtex4-1}

\usepackage{bm}
\usepackage{graphicx}
\usepackage{color}
\usepackage{amsmath}
\usepackage{hyperref}

\usepackage{changes}

\newcommand{\Ba}{Ba$^+$ }

\newcommand{\Rbplus}{Rb$^+$ }

\newcommand{\commentOut}[1]{}
\newcommand{\kb}{\textrm{k}_\textrm{B}}

\begin{document}

\title{Energy scaling of cold atom-atom-ion three-body recombination}

\author{Artjom Kr\"{u}kow}
\author{Amir Mohammadi}
\author{Arne H\"{a}rter}
\author{Johannes Hecker Denschlag}

\affiliation{Institut f\"{u}r Quantenmaterie and Center for Integrated Quantum Science and
Technology IQST, Universit\"{a}t Ulm, 89069 Ulm, Germany}

\author{ Jes\'{u}s P\'{e}rez-R\'{i}os}
\author{Chris H. Greene }

\affiliation{Department of Physics and Astronomy,
Purdue University,  47907 West Lafayette, IN, USA}

\date{\today}

\begin{abstract}
We study three-body recombination of \Ba + Rb + Rb in the mK regime where a single  $^{138}$Ba$^{+}$ ion in a Paul trap is immersed into a cloud of ultracold $^{87}$Rb atoms. We measure the energy dependence of the three-body rate coefficient $k_3$ and compare the results to the theoretical prediction, $k_3 \propto E_{\textrm{col}}^{-3/4}$ where $E_{\textrm{col}}$ is the collision energy.
We find agreement if we assume that the non-thermal ion energy distribution is determined by at least two different micro-motion induced energy scales. Furthermore, using classical trajectory calculations we predict how the median binding energy of the formed molecules scales with the collision energy.
Our studies give new insights into the kinetics of an ion immersed into an ultracold atom cloud and yield important prospects for atom-ion experiments targeting the s-wave regime.
\end{abstract}

\maketitle

When three atoms collide, a diatomic molecule can form in a three-body recombination (TBR) process.
In cold neutral atomic gases, TBR was investigated for spin-polarized hydrogen as well as alkalis (see e.g. \cite{Hess-1983, Burt-1997, Esry-1999}).
In the context of Bose-Einstein condensation, TBR plays a crucial role as a main loss mechanism.
By now, the scaling of TBR as a function of collision energy and scattering lengths in \textit{neutral} ultracold gases has been investigated in detail \cite{Esry-2005}.
When considering TBR in atom-ion systems, one can expect three-body interactions to be more pronounced due to the underlying longer-range $r^{-4}$ polarization potential.
Energy scaling of TBR in charged gases was studied at temperatures down to a few K, especially for hydrogen and helium due to their relevance in plasmas and astrophysics (e.g. \cite{Kristic2003,Plasil2012}).
Depending on the studied temperature range a variety of power laws was found but not a common threshold law.
The recent development of novel hybrid traps for both laser cooled atoms and ions has opened the possibility to investigate cold atom-ion interactions and chemical reactions in the mK-regime and below.
First experiments in such setups studied elastic and reactive two-body collisions (e.g. \cite{Grier-2009,Zipkes2010PRL,Schmid-2010,Hall-2011, Sullivan-2012,Ravi-2012, Smith-2012, Haze-2013}).
In accordance with the well-known Langevin theory, the corresponding reactive rates were measured to be  independent of the collision energy \cite{Zipkes2010PRL, Hall-2011}.
Very recently we predicted a theoretical threshold law on the scaling properties for cold atom-atom-ion three-body collisions \cite{JPR-2015}.
Understanding the scaling of reaction rates with quantities such as the collision energy is crucial for fundamentally understanding TBR and for the prospects of the experimental realization of ultracold s-wave atom-ion collisions.
Furthermore, as we will show here, studying TBR allows for insights into the kinetics of an ion immersed in a cloud of atoms.
Experimentally, TBR in the mK regime was recently observed for Rb$^+$ + Rb + Rb \cite{Harter-2012} and \Ba + Rb + Rb \cite{Kruekow-2016}.
In the \Ba experiments TBR was already dominating over two-body reactions even for moderate atomic densities of $10^{12}\:$cm$^{-3}$.

This letter reports on the combined theoretical and experimental investigation of the energy scaling of three-body atom-atom-ion collisions  in the  mK regime.
We measure the TBR rate coefficient $\overline{k}_3$ of \Ba in an ultracold Rb cloud as a function of the mean collision energy of the ion, $\overline{E}_{\textrm{col}}$, which we control via the excess micromotion (eMM) of the Paul trap.
 $\overline{k}_3$ is formally distinguished from $k_3$ which is the TBR rate coefficient for a precise collision energy $E_{\textrm{col}}$ in the center of mass frame.
By averaging $k_3$ over the ion energy distribution $\overline{k}_3$ is obtained.
We calculate $k_3$ using classical trajectory calculations (CTC) \cite{JPR-2014,JPR-2015} and derive its energy scaling,  $k_3 \propto E_{\textrm{col}}^{-3/4}$.
Agreement is found between theory and experiment if we assume that the energy distribution of the ion depends on multiple energy scales due to various sources of excess micromotion.
Besides the prediction of $k_3$, the CTC calculations  also provide the binding energy distribution of the formed molecules and the scaling properties of these distributions when the collision energy is varied.

The experiments are performed in a hybrid apparatus that has already been described in detail elsewhere \cite{Schmid2012}.
After loading a single $^{138}$Ba$^+$ ion by isotope selective, resonant two-photon ionization it is stored in a linear Paul trap driven at a frequency of $4.21\:\textrm{MHz}$ with radial and axial trapping frequencies of $(\omega_r; \omega_a) =2\pi \times (59.5;38.4)\:\textrm{kHz}$, respectively.
There, it is laser cooled to Doppler temperatures of $\approx 0.5\:$mK.
In order to perform our experiments in the electronic ground state we switch off the cooling and repumper light, before immersing the ion into the ultracold atomic cloud.

Once in the cloud, there is a complicated interplay of elastic two-body atom-ion collisions and the driven micromotion of the Paul trap.
This interplay leads to a non-Maxwell-Boltzmann distribution of the ion's kinetic energy $E_{\textrm{kin}}$ \cite{Devoe2009, Zipkes2010PRL, Cetina2012, Krych2013} with an equilibration time on the ms timescale \footnote{The equilibration time can be estimated from the Langevin collision rate which at our
given density is about $4\:$kHz.}.
The average kinetic energy $\overline{E}_{\textrm{kin}}$ of the ion in the atom cloud is then determined by the available energy sources for the ion, such as the eMM energy \cite{Zipkes2010PRL}.
In our experiment we can adjust $\overline{E}_{\textrm{kin}}$ by controlling one part of the eMM energy, $E_{\textrm{fMM}}$, which is set via static electric fields.
Concretely, we can write $\overline{E}_{\textrm{kin}}= c_{dyn}  (E_{\textrm{fMM}} + E_{\textrm{min}})$, where the offset energy $E_{\textrm{min}}$ contains all other energy contributions, \textit{e.g.} phase micromotion ($\phi$MM) \cite{Berkeland1998} or residual collisional effects \cite{Cetina2012, Krych2013}.
The proportionality factor $c_{dyn} \approx 5.0$, which depends on the atom-ion mass ratio and the trap parameters, is extracted from a MC calculation similar to \cite{Zipkes-2011}.
We can tune $E_{\textrm{fMM}}$ accurately between 5$\mu$K$\times k_B$ and 100mK$\times k_B$.
$E_{\textrm{min}}$, on the other hand, is not known precisely.
From independent measurements and MC calculations based on the scaling of elastic atom-ion collisions, we estimate $E_{\textrm{min}}$ to be in the range between 200 and 800 $\mu$K$\times k_B$.

The cloud consists of $N \approx 1.2 \times 10^5$ $^{87}$Rb atoms at a temperature of $T_\textrm{at} \approx 700\:$nK with a peak density of $n \approx 19 \times 10^{11}\:\textrm{cm}^{-3}$.
It is cigar shaped with a radial and axial size of roughly $10\:\mu$m and $50\: \mu$m respectively.
The atoms are spin polarized ($F=1, m_F=-1$) and confined in a far off-resonant crossed optical dipole trap at a wavelength of 1064$\:$nm with a trap depth of $\approx 10\:\mu$K$\times k_B$.
We shift the ion into the cloud over a distance of $120\: \mu$m within 2 ms by changing the endcap voltage of the linear ion trap.
After an interaction time of $\tau=300\:$ms, during which the Ba$^+$ ion is typically lost with a probability of up to 65 \%, we separate the two traps again and detect whether the Ba$^+$ ion is still present.
For this, we shine a laser cooling beam focused to a waist of $20\: \mu$m through the Paul trap center and collect the possible fluorescence on a EMCCD camera for 100$\:$ms.
If no \Ba is detected, we conclude that a reaction must have taken place during $\tau$
\footnote{
We note that our detection scheme cannot detect a reaction if the final product (e.g. after a secondary process) is again a cold Ba$^+$ ion.
From parallel experiments, however, we have no evidence for such a reaction outcome.}

\begin{figure}[t]
\includegraphics[width=\columnwidth]{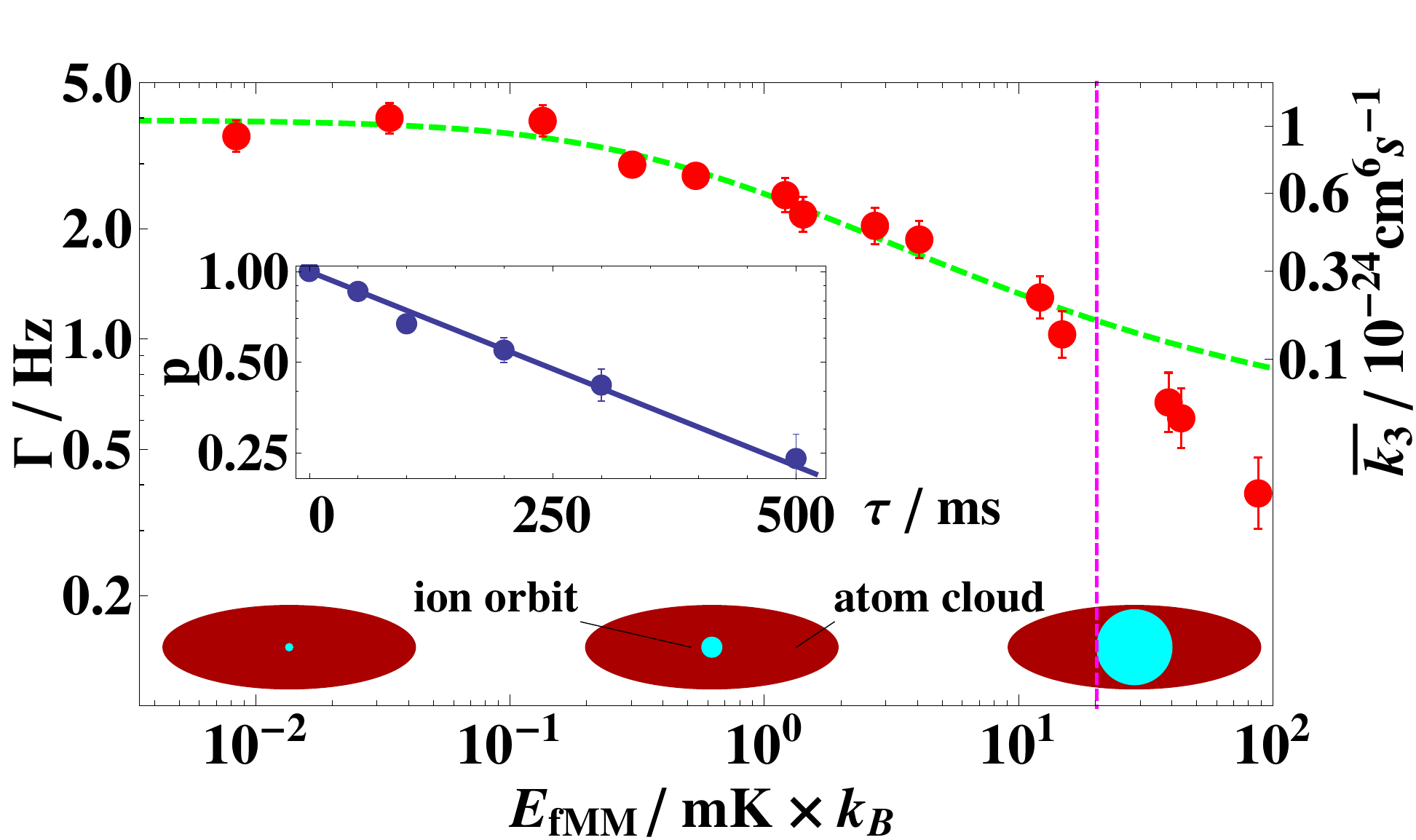}
\caption{{\bf(color online)} Double-logarithmic plot of the measured loss rate $\Gamma$ for \Ba as a function of the tuned eMM energy $E_{\textrm{fMM}}$.
Red circles are the experimental data, the curve represents a fit of Eq. (1) (see text for details). The corresponding values of $\overline{k}_3$ are indicated on the right hand side.
{\bf(inset)} Logarithmic plot of decay curve of the \Ba ion. $p$ is the probability to recover \Ba after interacting with Rb.
 The straight line is an  exponential fit to the data.
{\bf(sketch)} The sketch shows the ion orbit in the atom cloud. With increasing ion energy its orbit becomes comparable to the atom cloud size.
}
\label{fig1}
\end{figure}

Repeating the single ion experiment roughly 170 times we extract the probability $p$
that \Ba is still present.
For the given experimental settings the ion loss is well described by an exponential decay of the form $p = \exp (-\Gamma  \tau)$.
This can be seen in the inset of Fig. \ref{fig1}, where we plot $p$ as a function of interaction time $\tau$ measured at $E_{\textrm{fMM}}\approx 8\:\mu$K.
Fig. \ref{fig1} plots the loss rate $\Gamma$ as a function of $E_{\textrm{fMM}}$.
A Ba$^{+}$ ion in our experiment is lost either by a two-body charge transfer or by a three-body event \cite{Kruekow-2016}.
The corresponding loss rate $\Gamma$ of the ion is $\Gamma = -n k_{2}-n^{2} \overline{k}_3$.
The charge transfer rate coefficient $k_2$ has been previously measured for  \Ba + Rb, $ k_2=3.1(6)(6)\times 10^{-13}\textrm{cm}^{3}$/s (statistical and systematic errors in parentheses)  \cite{Kruekow-2016} (see also \cite{Schmid-2010, Hall-2013}), and contributes less than $1\:$Hz to the loss rate $\Gamma$ for the given atomic density.
Also, it  has been verified that  $k_2$ is energy independent \cite{Grier-2009,Zipkes2010PRL,Hall-2011},  in  consistency with Langevin theory. By subtracting this constant $k_2$-loss from  $\Gamma$ and dividing by the (constant) density $n^2$ we obtain $\overline{k}_3$ (see Figs. 1 and 3b).
Clearly, $\overline{k}_3$ is energy dependent.
As we will discuss later, we expect a scaling of $k_3$ with a power law, $k_3 \propto E_{\textrm{col}}^\alpha$.
Neglecting the atom motion due to ultracold temperatures we can express $E_{\textrm{col}}$ in terms of the ion kinetic energy $E_{\textrm{kin}}$, $E_{\textrm{col}}= (1-\frac{m_{\textrm{Ba}}}{m_{\textrm{Ba}}+2m_{\textrm{Rb}}}) E_{\textrm{kin}}$.
We attempt to describe the scaling of the measured $\overline{k}_3$ with a power law $\overline{k}_3 \propto \overline{E}_{\textrm{kin}}^{\alpha}$ by fitting the expression
\begin{equation}
\overline{k}_3= \overline{k}_{3,\textrm{min}}  \left[(E_{\textrm{fMM}}+E_{\textrm{min}})/E_{\textrm{min}}\right]^{\alpha},	
\label{eqn_k3}
\end{equation}
to the data.
Here, $E_{\textrm{min}}$ and $\alpha$ are free parameters.
The constant $\overline{k}_{3,\textrm{min}}=1.04(4)(45)\times 10^{-24}\textrm{cm}^{6}$/s is $\overline{k}_3$  at $E_{\textrm{fMM}} = 0$ and was determined in a parallel study \cite{Kruekow-2016}.
For the fit we discard data points above $E_{\textrm{fMM}}>20\: \textrm{mK} \times \kb$, as for such energies, the ion is not localized well enough in the center of the cloud.
It probes areas of the atomic cloud at lower densities, hence decreasing the observed loss rate (see sketch in Fig. 1).
The fit yields $\alpha=-0.46(9)$ and $E_{\textrm{min}}=410(180) \: \mu$K$\times \textrm{k}_\textrm{B}$ (green dashed line in Fig. \ref{fig1}), with the errors denoting a 1$\sigma$ statistical uncertainty of the fitted values.
Interestingly, in our previous study of TBR of  \Rbplus + Rb + Rb \cite{Harter-2012} we observed a similar scaling exponent of $\alpha=-0.43$.

 \begin{figure}[t]
\centering
 \includegraphics[width=\columnwidth]{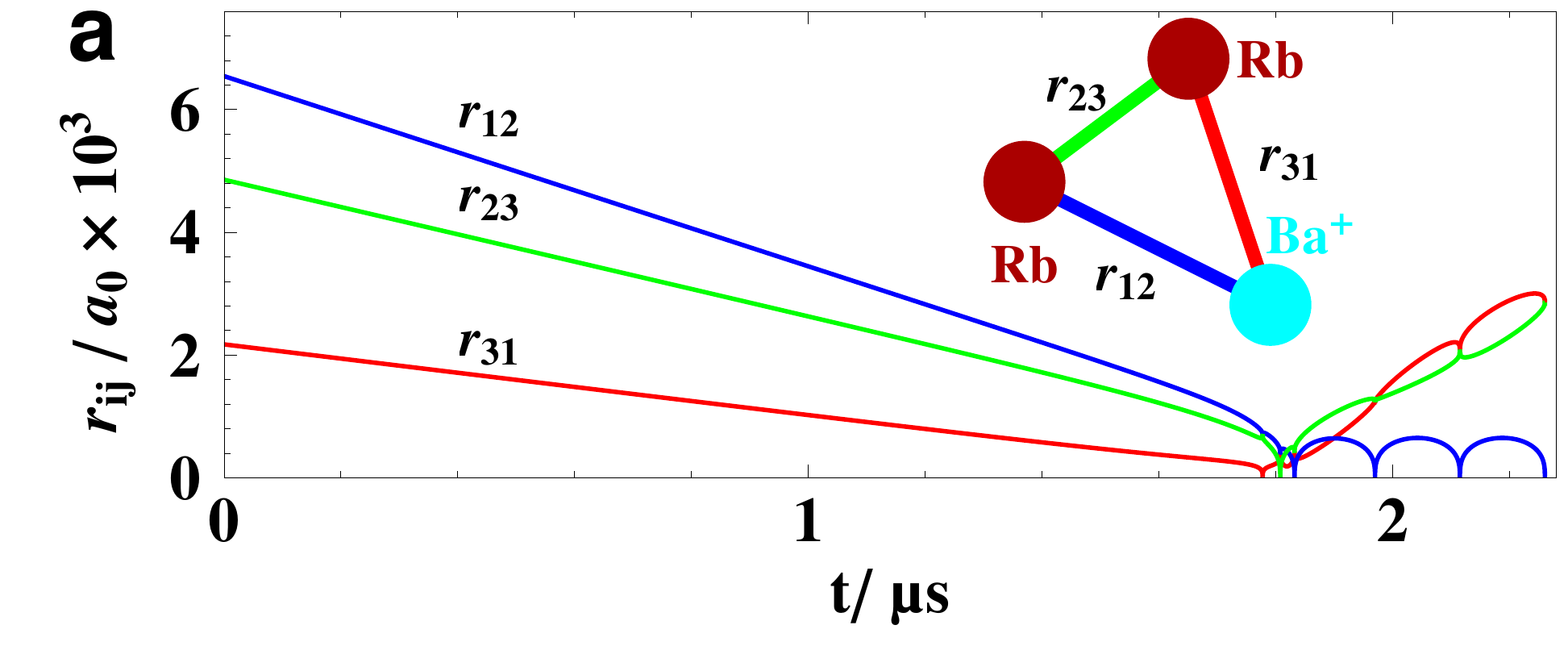}
 \includegraphics[width=\columnwidth]{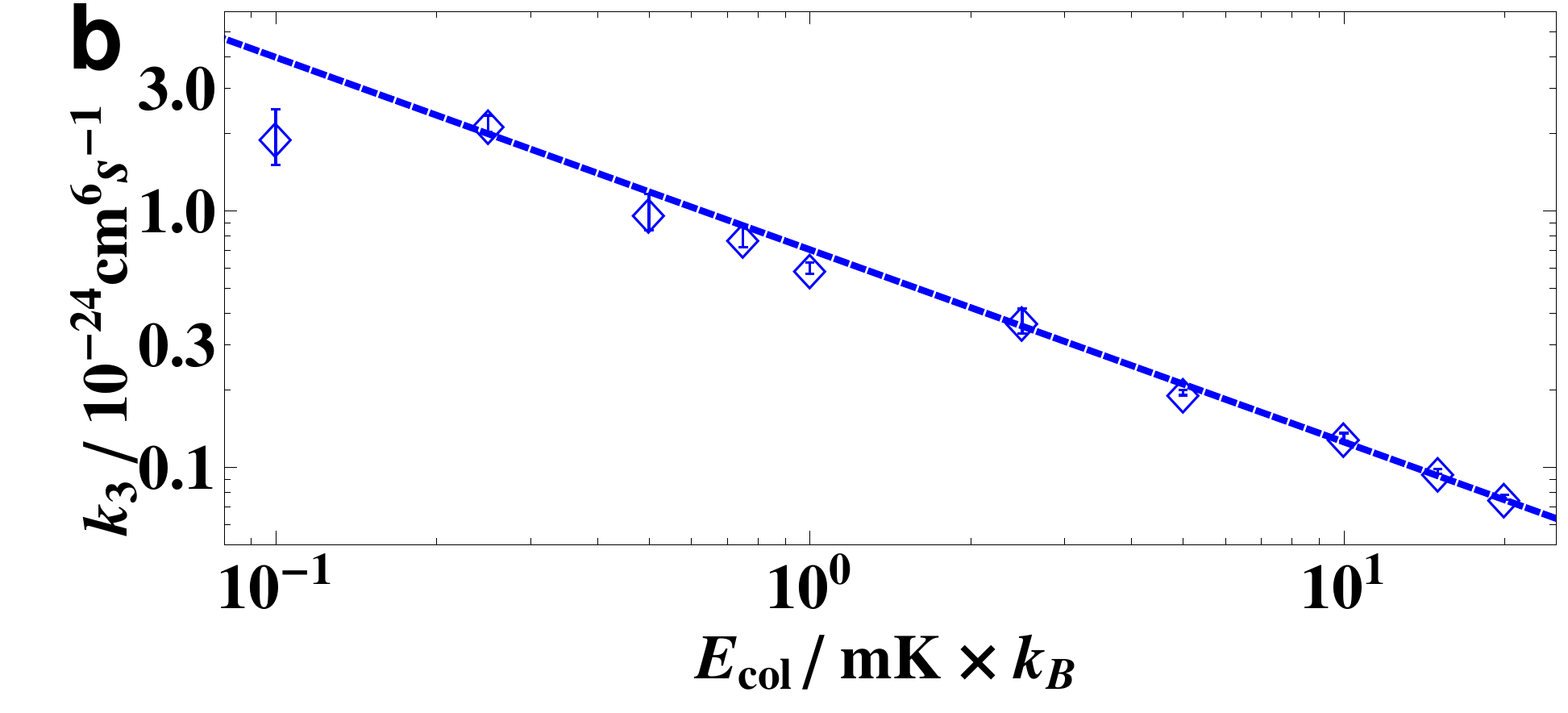}
 \caption{
\textbf{(color online)} \textbf{(a)} A typical trajectory at a collision energy of 100$\:\mu$K$\times \kb$ associated with the three-body collision Ba$^{+}$ + Rb + Rb that leads to the formation of BaRb$^{+}$.
Shown the distances $r_{ij}$ between the particles as indicated in the sketch.
\textbf{(b)} Double log plot of $k_3$ obtained with CTC for Ba$^{+}$ + Rb + Rb  as a function of the collision energy $E_{\textrm{col}}$ (circles). The straight line shows the analytically predicted power law dependence $k_{3}\propto E_{\textrm{col}}^{-3/4}$.
}
\label{fig2}
\end{figure}

We now turn to investigate the scaling of TBR theoretically with a classical trajectory calculation (CTC) formalism.
A classical treatment of the collision dynamics is appropriate, since the experiments described here in general involve much higher energies than the threshold energy of $\sim 50\:\textrm{nK} \times \kb$ for entering the s-wave regime of Ba$^{+}$-Rb.
We have adapted a recently developed method for the calculation of three-body recombination cross sections based on classical trajectories  \cite{JPR-2014,JPR-2015} for the study of atom-atom-ion recombination.
The method employed relies on mapping the three-body problem into a 6-dimensional configuration space, described in hyperspherical coordinates, after separating out the center of mass motion \cite{JPR-2014}.
Since the kinetic energy of the ion is typically several orders of magnitude higher than the temperature of the ultracold neutral atoms we fix one of the hyperangles associated to the ratio of the atom-ion versus the atom-atom initial momentum, guaranteeing that in the center of mass coordinate system 95 \% of the collision energy $E_{\textrm{col}}$ is along the direction of the ion.
In the classical trajectory calculations only  Rb-Rb collisions in triplet states are considered and spin flip transitions are neglected.
For the Rb-Rb pair interaction we employ the $a^{3}\Sigma_u^+$ potential of Strauss {\it et al.}  \cite{Strauss-2010}.
On the other hand, the Ba$^{+}$-Rb interaction potential is taken to be $-C_{4}(1-(r_{m}/r)^{4}/2)/r^{4}$, where $C_4=$ 140 a.u. denotes the experimental long-range value of the interaction and $r_{m}$ represents the position of the minimum of the potential, taken from Ref. \cite{Krych-2011}.

The TBR rate for Ba$^{+}$ + Rb + Rb  has been computed by running $10^{5}$ trajectories per collision energy.
We checked that during the simulation the total energy and angular momentum is conserved up to the fifth decimal place.
Details about the numerical method employed to solve Hamilton's equations of motion, in conjunction with the sampling of the initial conditions, can be found in \cite{JPR-2014}.
Fig. \ref{fig2}a shows a three-body trajectory that results in a recombination event with a collision energy of 100 $\mu$K$\times \textrm{k}_\textrm{B}$.
This particular trajectory leads to large size ($\sim$ 800 a$_{0}$), very weakly-bound  molecular ion.
Counting the fraction of trajectories that lead to molecule formation we can extract the TBR rate coefficient $k_3$ for Ba$^{+}$ + Rb + Rb. Fig. \ref{fig2}b plots $k_3$ as a function of collision energy $E_{\textrm{col}}$.
We compare these CTC calculations (diamonds) with an analytically derived scaling law \cite{JPR-2015} where $k_{3} \propto E_{\textrm{col}}^{-3/4}$ (dashed line in Fig. \ref{fig2}b and find very good agreement.
\begin{figure}[t]
\includegraphics[width=\columnwidth]{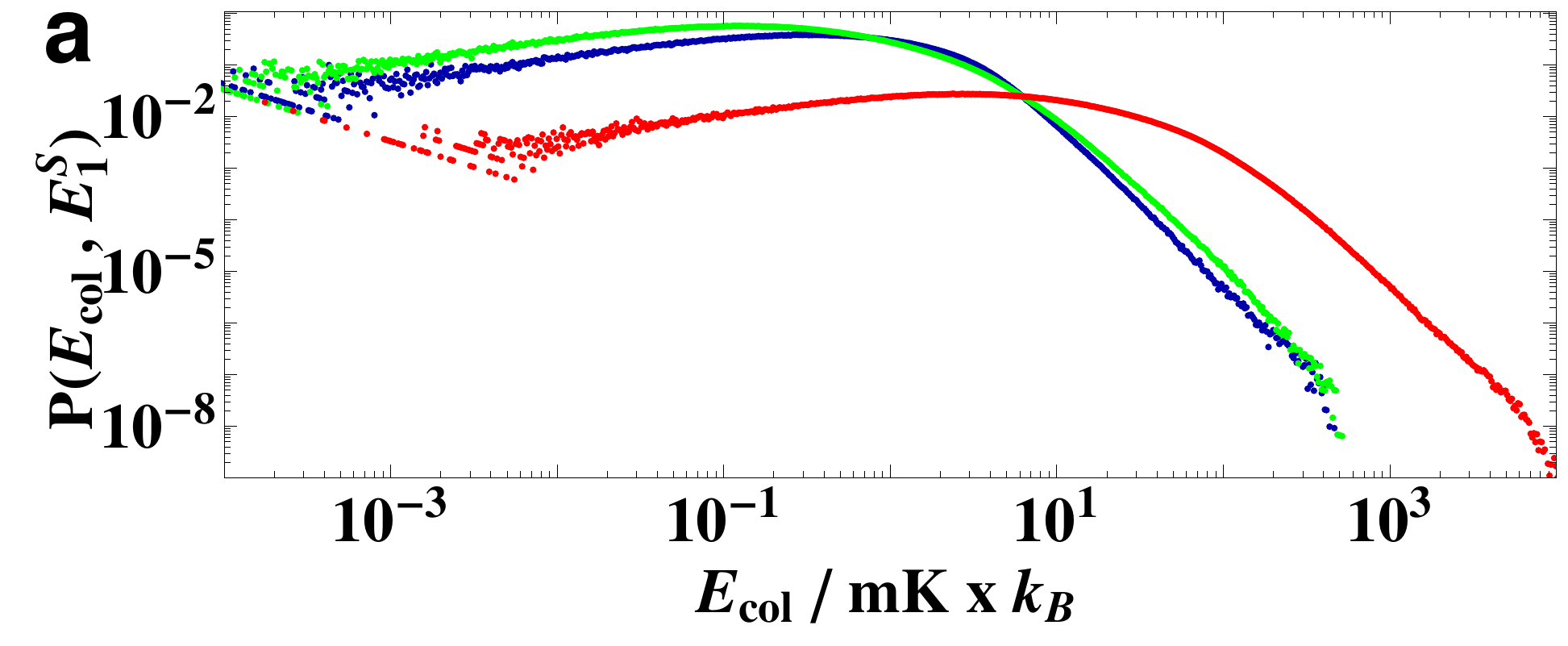}
\includegraphics[width=\columnwidth]{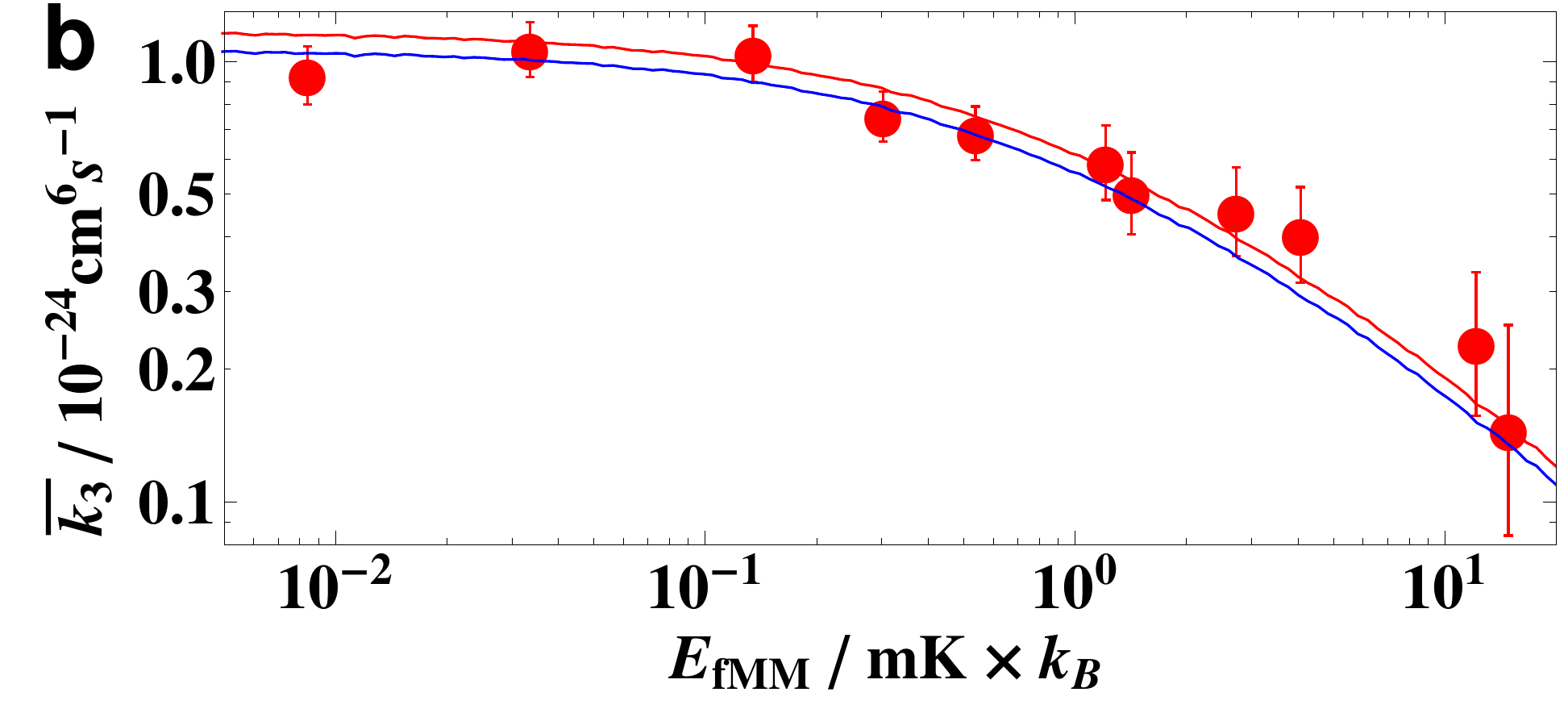}
\caption{
\textbf{(color online)}
\textbf{(a)} Calculated ion energy distributions $P(E_{\textrm{col}}, E^S_1 )$
each with a  single energy scale $E^S_1$.
An energy of $E^S_1=E_{\textrm{fMM}}=1\:$mK ($20\:$mK) was used for the green (red) distribution.
Choosing $E^S_1=E_{\phi \textrm{MM}}=1\:$mK produces the blue distribution, which has a different shape compared to both previous distributions.
\textbf{(b)} Comparison of the experimental (full circles) $\overline{k}_3$ data as a function of $E_{\textrm{fMM}}$ with the full calculation (blue line). 
The red line is the same calculation but multiplied by 1.1.
}

\label{fig3}
\end{figure}

Strikingly, the theory prediction of $\alpha = -0.75$ does not seem to agree well with the experimentally observed value of $\alpha =-0.46(9)$ from the fit of Eq. \ref{eqn_k3} to our data.
We explain this discrepancy as follows.
In contrast to the theoretical approach where $k_3$ is determined for a precisely defined collision energy $E_{\textrm{col}}$,  in the experiments we observe $\overline{k}_3$, an average over a distribution  $P({E_{\textrm{col}}}, \{ E^S_i\})$ of collision energies, calculated as
\begin{equation}
\label{eqn-k3int}
\overline{k}_3 (\{ E^S_i\}) =  \int k_{3}(E_{\textrm{col}}) \ P(E_{\textrm{col}}, \{ E^S_i\}) \ dE_{\textrm{col}}.
\end{equation}
Here, $\{ E^S_i\}$ is a list of the relevant energy scales that determine the distribution, such as the experimentally tuned $E_{\textrm{fMM}}$ or $E_{\phi \textrm{MM}}$.
We extract these distributions with a MC calculation based on \cite{Zipkes-2011}.
If only a single scale $E^S_1$ is present, the energy distributions can be expressed as functions of the ratio $E_{\textrm{col}}/ E^S_1$,
\begin{equation}
\label{eqn-rescale}
P(E_{\textrm{col}}, E^S_1) dE_{\textrm{col}}= \tilde{P}(E_{\textrm{col}}/ E^S_1) dE_{\textrm{col}}/ E^S_1.
\end{equation}
Fig. \ref{fig3}a shows three calculated distributions each with its own scale $E^S_1$.
The distributions $P(E_{\textrm{col}}, E_{\textrm{fMM}})$ for $E_{\textrm{fMM}}$ = 1$\:$mK (green) and 20$\:$mK (red) have the exact same shape, a consequence of Eq. \ref{eqn-rescale}.
The third distribution $P(E_{\textrm{col}}, E_{\phi \textrm{MM}}=1\:$mK$)$ (blue), generated with a phase micromotion has a somewhat different shape.
Using Eq. \ref{eqn-k3int} one can show that distributions which satisfy Eq. \ref{eqn-rescale} translate the power law $k_3 \propto E_{\textrm{col}}^{-3/4}$ into $\overline{k}_3 \propto (E^S_1) ^{-3/4}$.
In our experiment, however, where at least two energy scales, $E_{\textrm{fMM}}$ and $E_{\textrm{min}}$ occur, this translation of the scaling breaks down and Eq. \ref{eqn_k3} cannot be used in the data analysis anymore.
Instead, we calculate $\overline{k}_3$ with Eq. \ref{eqn-k3int} to directly compare theory and experiment.
The choice and magnitude of $E_{\textrm{min}}$ is the only free model parameter.
Here, we assume that $E_{\textrm{min}}$ is entirely determined by phase micromotion, $E_{\textrm{min}} = E_{\phi \textrm{MM}}$. The phase micromotion is chosen to be shared equally between both pairs of opposing radio frequency (RF) driven electrodes \cite{Berkeland1998}. 
Fig. \ref{fig3}b shows the experimental $\overline{k}_3$ (full circles), together with the calculation (blue solid line) with $E_{\phi \textrm{MM}}=790\:\mu$K,
\footnote{Such a phase micromotion can be caused by a length difference of $\Delta l \approx 3\:$mm between the cables supplying opposing RF electrodes, which is well within the tolerances of our setup.}.
The shape of the theory curve describes the experimental data quite well, apart from an overall factor of about 1.1 (see blue and red solid lines). 
In general, the overall magnitude and energy dependence of $\overline{k}_3$ is reproduced by the presented {\it ab initio} CTC treatment down to the mK-Regime, which is remarkable as $E_{\textrm{min}}$ is the only free parameter.

\begin{figure}[t]
\centering
 \includegraphics[width=\columnwidth]{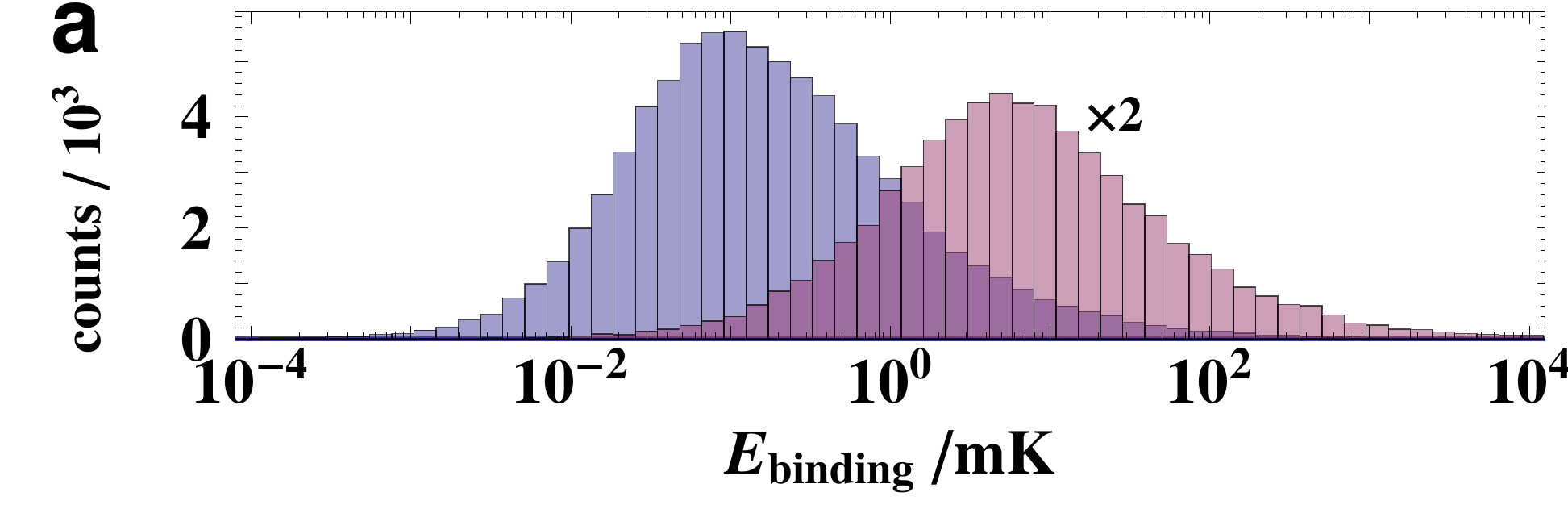}
 \includegraphics[width=\columnwidth]{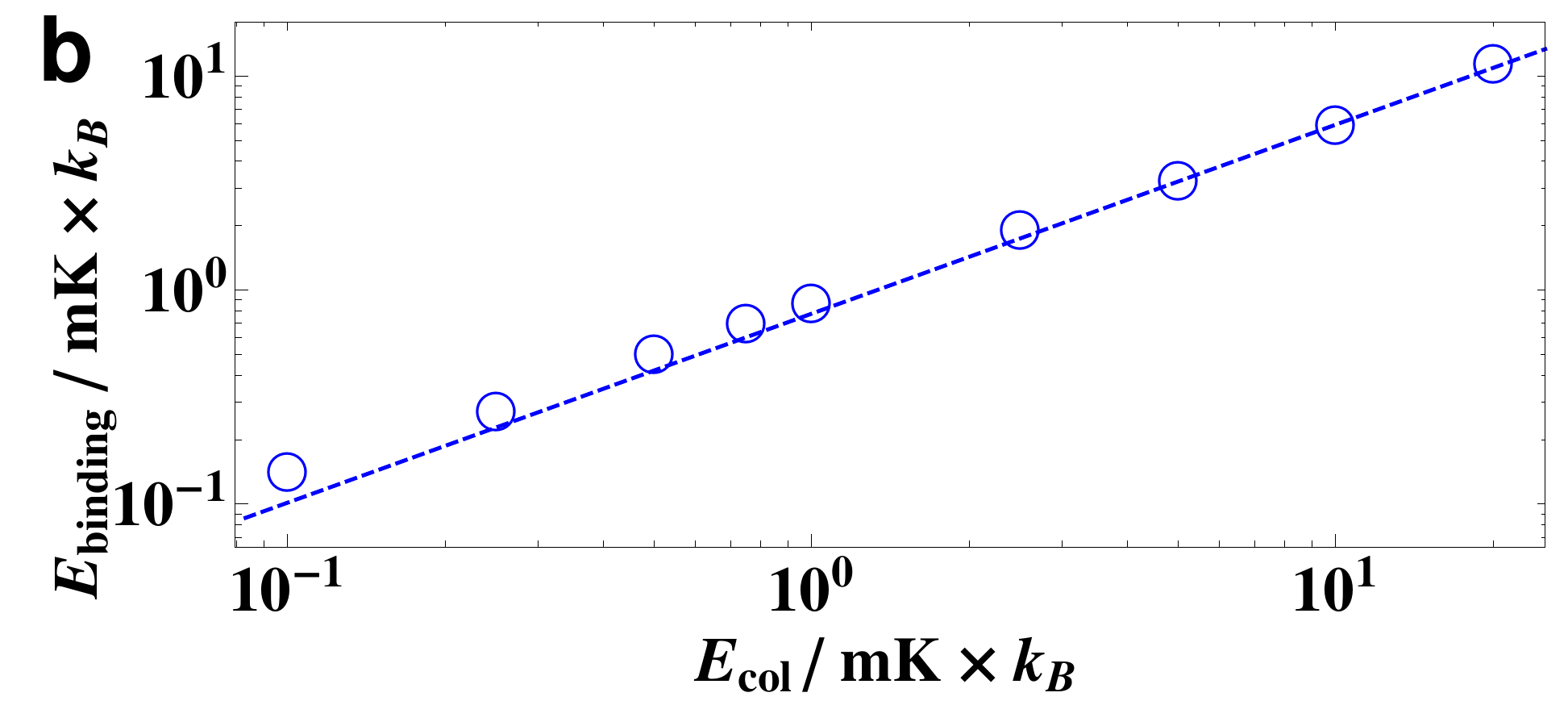}
  \caption{(color online)
\textbf{(a)} Logarithmically binned histogram of the binding energies at collision energies of 100$\:\mu$K$\times \textrm{k}_\textrm{B}$ (blue) and 10 mK$\times \textrm{k}_\textrm{B}$ (red). The second histogram is magnified by a factor of two.
\textbf{(b)} Double-logarithmic plot of the typical binding energy of the formed molecule as a function of the collision energy. The dashed line represents a power law fit.}
\label{fig4}
\end{figure}

We now turn to briefly discuss the molecular products after TBR.
In a previous study of TBR for He, it was suggested that the binding energy of the products is correlated with the collision energy \cite{JPR-2014}.
We find again the same behavior for TBR of an ion with two atoms.
Fig. \ref{fig4}a shows two logarithmically binned histograms of molecular binding energies after TBR.
The maximum of each histogram can be considered the typical binding energy and is shown in Fig. \ref{fig4}b as a function of the collision energy $E_{\textrm{col}}$.
A fit to a power-law dependence gives $E_{\textrm{binding}} \sim E_{\textrm{col}}^{0.88\pm0.02}$ for the energy range investigated here.
Thus our calculations suggest that the formation of deeply bound molecules after TBR  should be highly improbable at low collision energies.

The present CTC results also suggest that BaRb$^+$ should be the dominant product state of the three-body recombination in the collision energy range considered here.
Indeed, we have observed the formation of  BaRb$^+$ ions in our experiment.
However, collisional or light induced secondary processes lead to short lifetimes.
A detailed study of the initial TBR products and involved secondary reactions is currently in progress and needs to be discussed elsewhere.

In conclusion, we have investigated the energy scaling of three-body recombination in an atom-ion system down to mK energies.
Single \Ba ions in contact with ultracold Rb atoms have been used to measure the TBR rate coefficient $\overline{k}_3$.
Utilizing classical trajectory calculations, we numerically accessed the TBR rate coefficient $k_3$ for the \Ba + Rb + Rb  system for various collision energies.
We find a power law scaling of the form $k_3(E_{\textrm{col}}) \propto E_{\textrm{col}}^{\alpha}$ with an exponent  $\alpha = -3/4$.
Our experimental and theoretical studies indicate that the presence of several energy scales gives rise to energy distributions of the immersed ion which impede a direct application of scaling laws to the measured data.
The obtained energy scaling provides an important insight for prospects of atom-ion experiments in the ultracold regime, as the already strong TBR rate observed here will increase by another three orders of magnitude once the s-wave regime at $50\:$nK is reached.

This work was supported by the German Research Foundation DFG within the SFB/TRR21 and by the Department of Energy, Office of Science, under Award Number DE-SC0010545. A.K. acknowledges support from the Carl Zeiss Foundation. J.P.-R. and C.H.G. thank Francis Robicheaux for many fruitful discussions.
J.H.D. and C.H.G acknowledge inspiring interactions within program INT-14-1.

\end{document}